\documentclass[aps,pre,twocolumn,longbibliography,reprint]{revtex4-2}

\usepackage{graphicx}
\usepackage{amsmath}
\usepackage{subcaption}
\usepackage{enumitem}
\usepackage[colorlinks=true,hyperfootnotes=true,breaklinks=true]{hyperref}
\usepackage{cleveref}

\DeclareMathOperator{\Tr}{\mathrm{Tr}}

\begin{document}

\title{Heider balance on Archimedean lattices and cliques}

\author{Krzysztof Malarz}
%% \email{malarz@agh.edu.pl}
\thanks{ORCID~\href{https://orcid.org/0000-0001-9980-0363}{0000-0001-9980-0363}}

\author{Maciej Wo{\l}oszyn}
\email{woloszyn@agh.edu.pl}
\thanks{ORCID~\href{https://orcid.org/0000-0001-9896-1018}{0000-0001-9896-1018}}

\author{Krzysztof Ku{\l}akowski}
\thanks{ORCID~\href{https://orcid.org/0000-0003-1168-7883}{0000-0003-1168-7883}}

\affiliation{AGH University,
Faculty of Physics and Applied Computer Science,
al. Mickiewicza 30, 30-059 Krak\'ow, Poland}

\date{\today}

\begin{abstract}
We investigate the work function $U(T)$ for the Heider balance, driven by a thermal noise $T$,  on several planar networks that contain separated triangles, pairs of triangles, chains of triangles and complex structures of triangles. In simulations, the heat-bath algorithm is applied. Two schemes of link values updating are considered: synchronous and asynchronous (sequential). The latter results are compared with analytical calculations for small cliques. We argue that the actual shape of $U(T)$ is a consequence of a local topology rather than of a macroscopic ordering. Finally, we present the mathematical proof that for any planar lattice, perfect structural (Heider) balance is unreachable at $T>0$.
\end{abstract}

\maketitle

%% ###############################################
%% ###############################################
\section{Introduction}
%% ###############################################
%% ###############################################

The phenomenon of Heider (structural) balance is known for more than 70 years \cite{Heider,Bonacich}. It attracts continuous attention of
numerous computational scholars, as it seems to be an example of a macroscopic ordering which emerges as
a consequence of local interactions. From a sociological perspective, the problem deals with a cognitive dissonance, which appears when the partition of our acquaintances into enemies and friends remains fuzzy. From a computational point of view, the problem is to evaluate the stationary departure from a balanced state in the presence of noise. Such an algorithm could be used to predict conflicts, and perhaps to prevent them.

The sociophysical problem of structural balance \cite{Heider,Harary_1953,Cartwright_1956,Harary_1959,Davis_1967,Harary,Bonacich}
governed by Heider's algebra \cite{Heider}: 
\begin{itemize}[noitemsep]
\item a friend of my friend is my friend;
\item a friend of my enemy is my enemy;
\item an enemy of my friend is my enemy;
\item an enemy of my enemy is my friend
\end{itemize}
attracted attention of physicists (see References~\onlinecite{ISI:000247470200019,Belaza_2017} for recent reviews) as a three body problem based on interaction in triads of actors.

%% ---------------------------------------------------------------
\begin{figure}[htbp]
\begin{subfigure}[b]{.24\columnwidth}
\caption{\label{fig:tr-s+3}}
\includegraphics[width=0.99\textwidth]{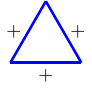}
\end{subfigure}
%% ---------------------------------------------------------------
\begin{subfigure}[b]{.24\columnwidth}
\caption{\label{fig:tr-s+1}}
\includegraphics[width=0.99\textwidth]{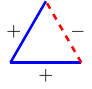}
\end{subfigure}
%% ---------------------------------------------------------------
\begin{subfigure}[b]{.24\columnwidth}
\caption{\label{fig:tr-s-1}}
\includegraphics[width=0.99\textwidth]{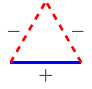}
\end{subfigure}
%% ---------------------------------------------------------------
\begin{subfigure}[b]{.24\columnwidth}
\caption{\label{fig:tr-s-3}}
\includegraphics[width=0.99\textwidth]{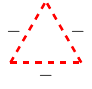}
\end{subfigure}
%% ---------------------------------------------------------------
\caption{\label{triangle}Heider's triads corresponding to balanced (\subref{fig:tr-s+3}~and~\subref{fig:tr-s-1}) and imbalanced (\subref{fig:tr-s+1}~and~\subref{fig:tr-s-3}) states. Solid blue lines and dashed red lines represent friendly ($+1$) and hostile ($-1$) relations, respectively}
\end{figure}
%% ---------------------------------------------------------------

When relations in every triangle obey Heider's rules mentioned above, the system is termed as balanced in Heider's sense---which means that only triangles presented in \Cref{fig:tr-s+3,fig:tr-s-1} are observed in the system.
Measures of structural balance based on the concepts of weak and strong balance with application to several real-world signed networks are discussed in Reference~\onlinecite{Kirkley2019}, and further statistical perspective used to assess the empirical unbalanced patterns is provided in~Reference~\onlinecite{Gallo2024} in particular for exponential random graphs.

The earlier studies have been devoted to triangular \cite{2007.02128}, diluted \cite{2005.11402} and densified triangulations \cite{2106.03054}; chains of nodes \cite{2008.06362}; classical random graphs \cite{2206.14226}; and complete graphs \cite{1911.13048,2009.10136}.  Several works deal with a balance-imbalance phase transition at finite temperature \cite{2009.10136,Kargaran_2020,2008.00537,2010.10036,2106.03054,PhysRevE.105.054105,Mohandas_2024}. In these works, numerical simulations are often backed by the mean-field approach \cite{2009.10136,2008.00537,PhysRevE.105.054105}. The balance-imbalance phase transition is found to be of first order (e.g. in Erd{\H o}s--R\'enyi networks \cite{Masoumi_2022}). In Reference~\onlinecite{Kargaran_2020}, the authors have indicated that as the usual form of the work function consists of a sum over particular triads, the triads are independent even if they share a link. Below we analyze how this assertion is influenced by a local topology of the network.

\begin{figure*}[htbp]
\begin{subfigure}[b]{.18\textwidth}
\caption{\label{fig:pic-AL-333333}}
\includegraphics[width=.99\textwidth]{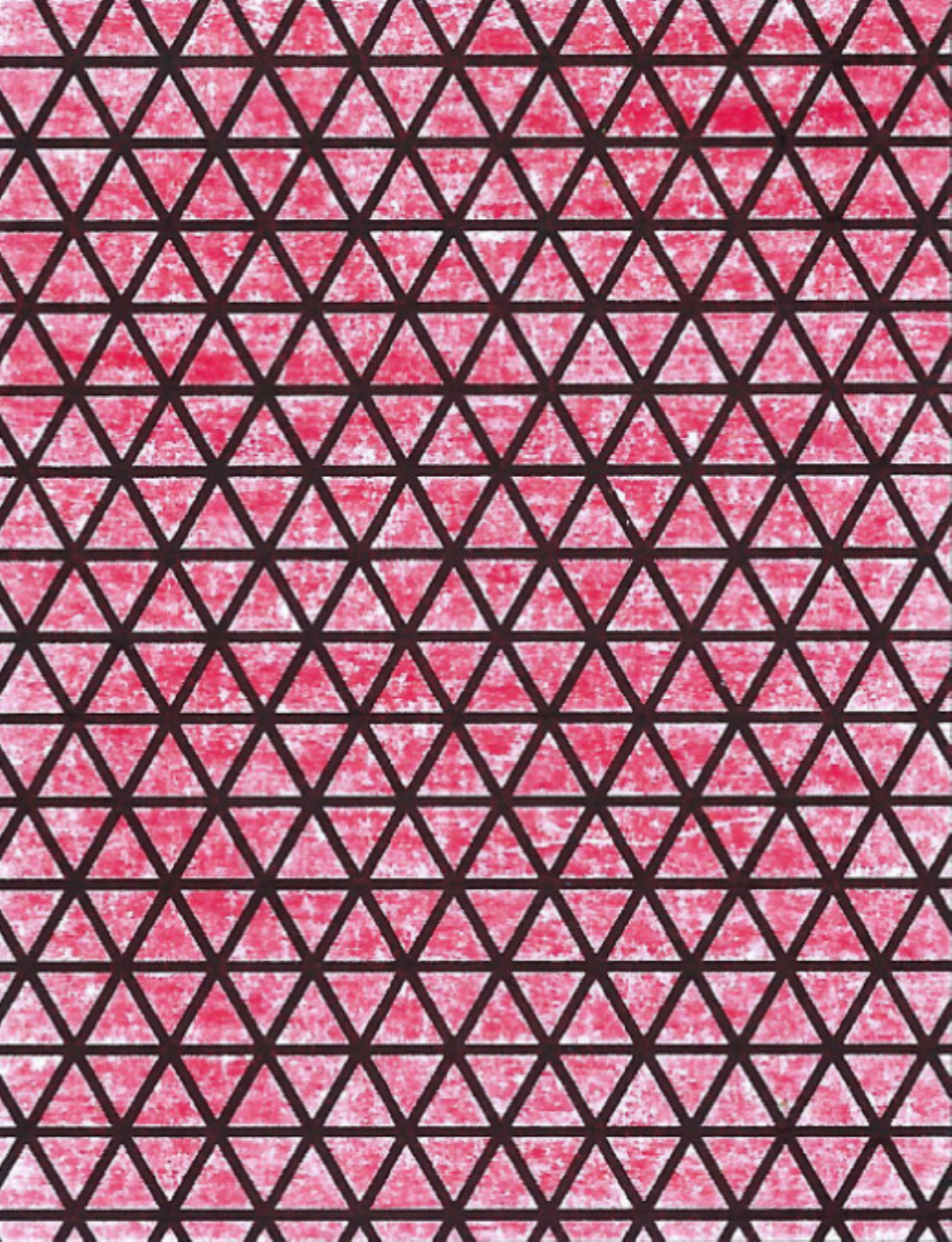}
\end{subfigure}
\hfill %% ---------------------------------------------------------------
\begin{subfigure}[b]{.18\textwidth}
\caption{\label{fig:AL-33434}}
\includegraphics[width=.99\textwidth]{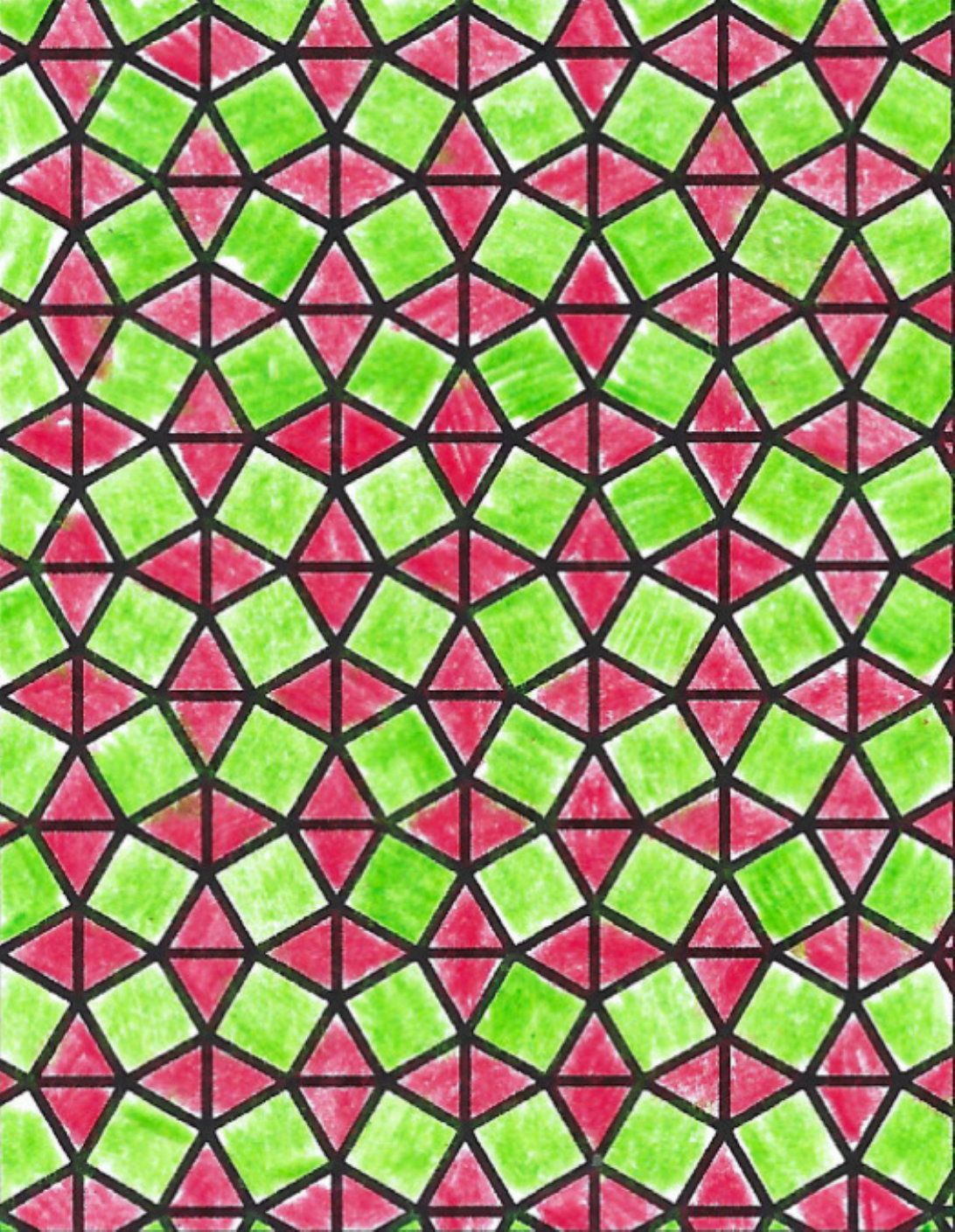}
\end{subfigure}
\hfill %% ---------------------------------------------------------------
\begin{subfigure}[b]{.18\textwidth}
\caption{\label{fig:AL-33344}}
\includegraphics[width=.99\textwidth]{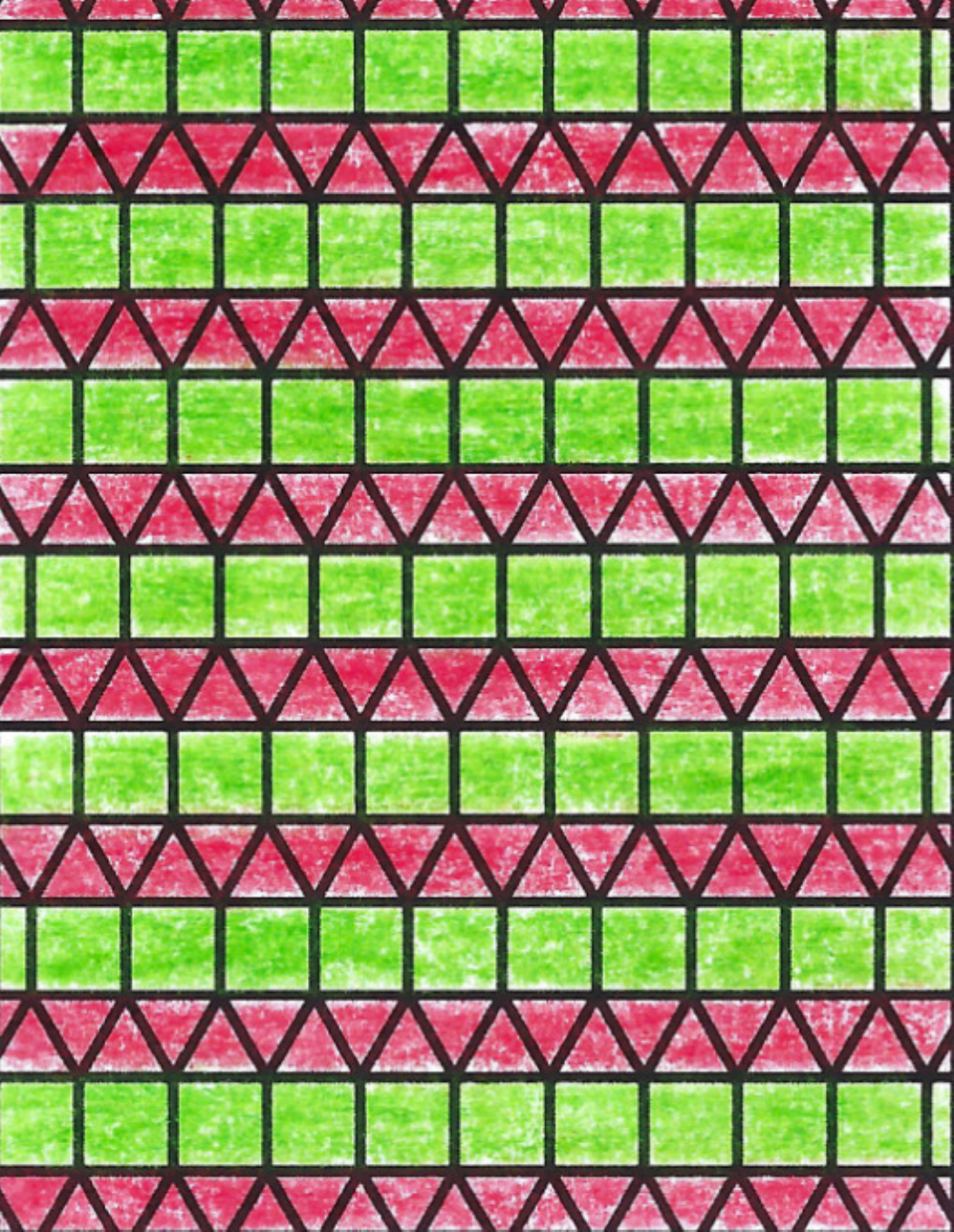}
\end{subfigure}
\hfill %% ---------------------------------------------------------------
\begin{subfigure}[b]{.18\textwidth}
\caption{\label{fig:AL-33336}}
\includegraphics[width=.99\textwidth]{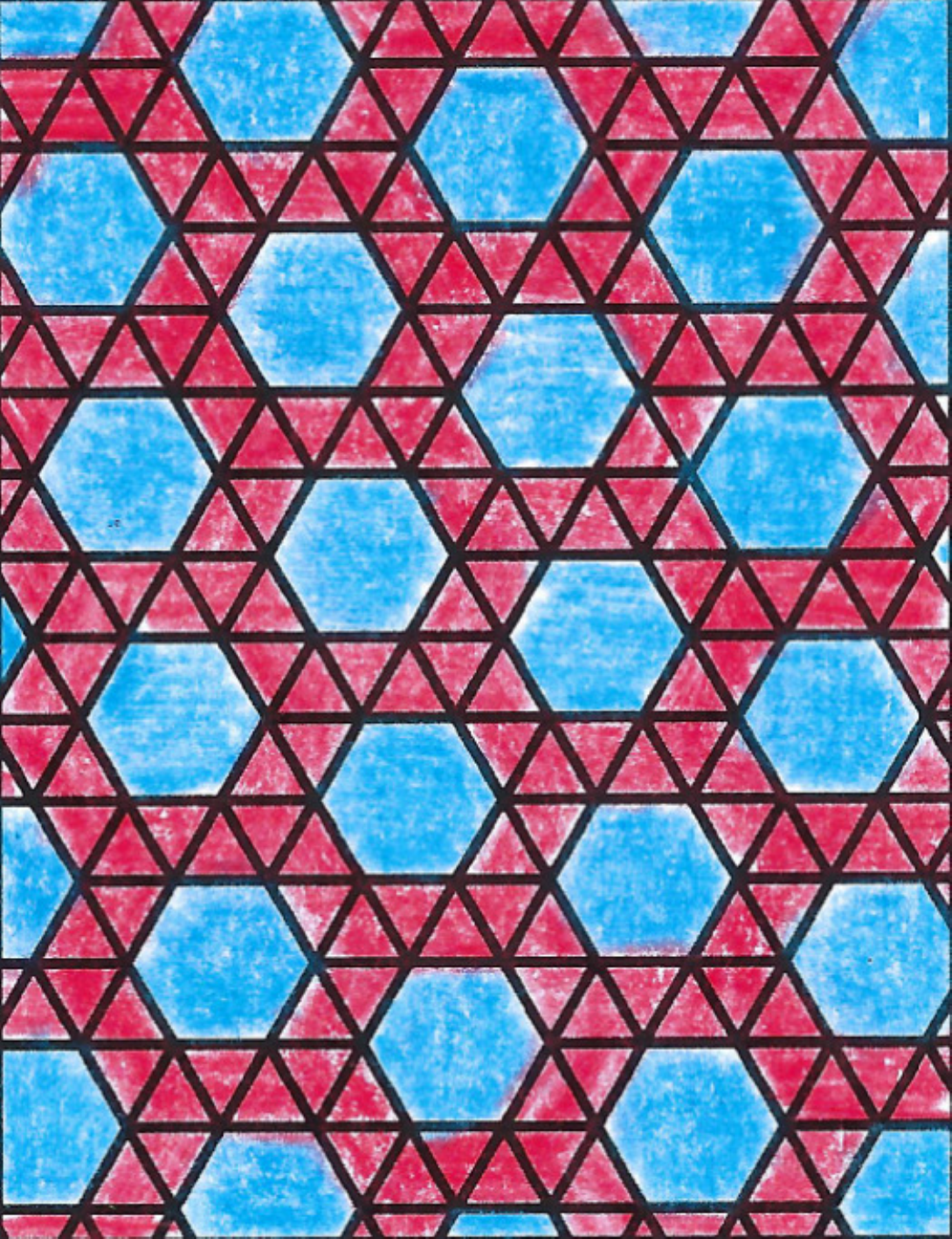}
\end{subfigure}
\hfill %% ---------------------------------------------------------------
\begin{subfigure}[b]{.18\textwidth}
\caption{\label{fig:pic-AL-3636}}
\includegraphics[width=.99\textwidth]{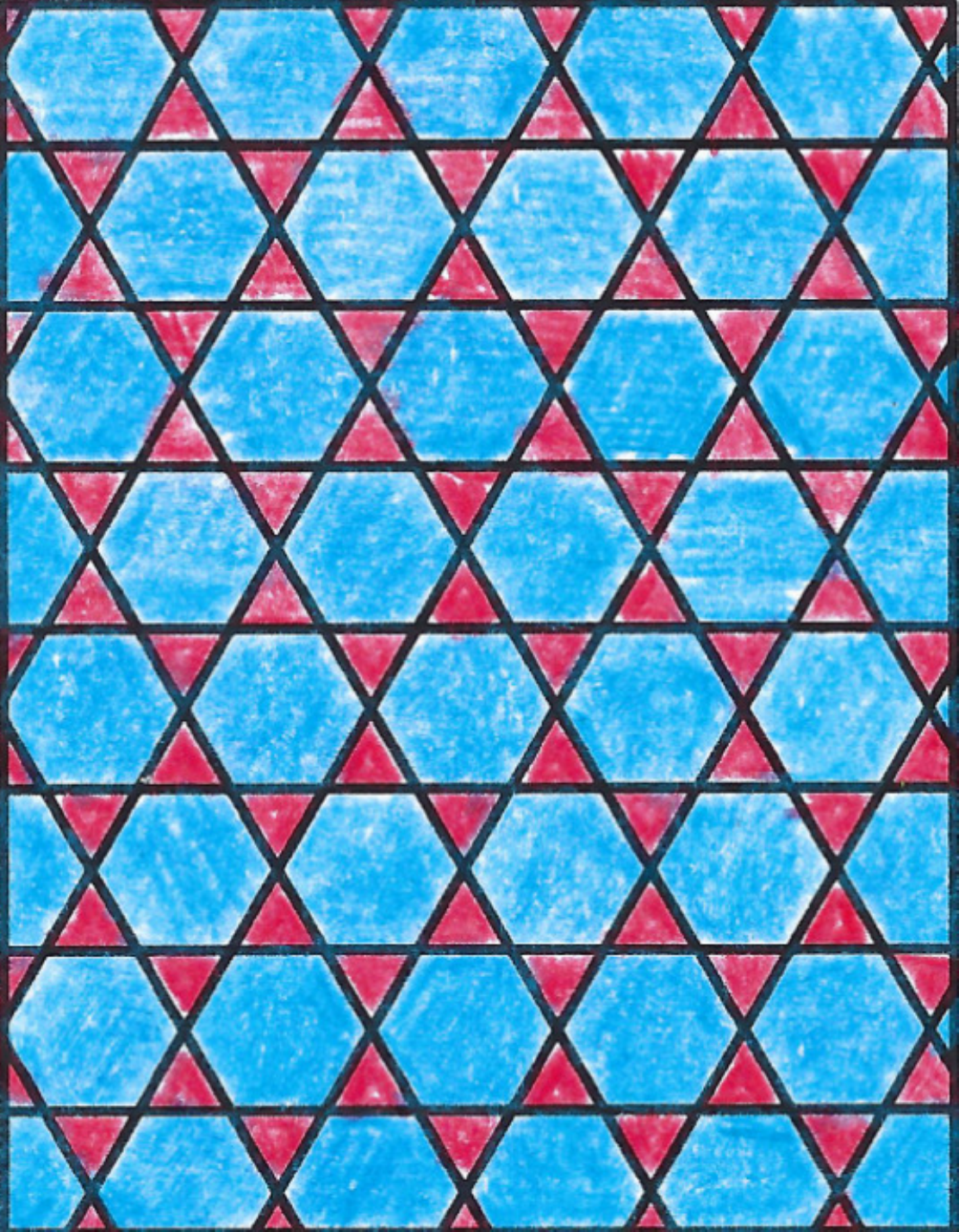}
\end{subfigure}
%% ---------------------------------------------------------------
\caption{\label{fig:AL}Shapes of the Archimedean lattices:
(\subref{fig:pic-AL-333333}) $(3^6)$;
(\subref{fig:AL-33434}) $(3^2,4,3,4)$;
(\subref{fig:AL-33344}) $(3^3,4^2)$;
(\subref{fig:AL-33336}) $(3^4,6)$;
(\subref{fig:pic-AL-3636}) $(3,6,3,6)$}
\end{figure*}

Recently, reaching structural balance has been investigated with thermal noise modeled either with heat-bath \cite{PhysRevE.99.062302,1911.13048,2009.10136} or Glauber dynamics \cite{PhysRevE.100.022303,2011.07501}. 
(A more refined dynamics based on the indirect reciprocity mechanism on a complete graph has been studied in Reference~\onlinecite{2404.15664}). In this paper, we continue studies of the influence of thermal noise on the possibility of reaching structural balance on Archimedean lattices, all of them planar (see \Cref{fig:AL}).
The results of the simulations are accompanied by analytical calculations.
The latter are also performed for cliques of four and five nodes.

Properties of triads of actors is a classic chapter of sociology since times of Georg Simmel \cite{Yoon_2013}. The Archimedean lattices, as being periodic structures composed of triads, are convenient as a medium for computational research. There are eleven Archimedean lattices \cite{Kepler_1619}, which contain only regular polygons allowing for covering plane with tiles and keeping the translational symmetry of the lattice.
The systematic names of these lattices come from the number of edges in tiles attached to every lattice node sorted in the possibly lowest lexicographic order---for example, we call lattice presented in \Cref{fig:AL-33434} $(3^2,4,3,4)$ and not $(4,3^2,4,3)$ or $(4,3,4,3^2)$.
In this way, the square lattice is called $(4^4)$, the triangular lattice is called $(3^6)$, the honeycomb lattice is called $(6^3)$ and kagom\'e lattice is called $(3,6,3,6)$.

%% ---------------------------------------------------------------
%% ---------------------------------------------------------------

The Archimedean lattices investigated here are:
\begin{itemize}
\item separated triangles: kagom\'e---$(3,6,3,6)$ lattice, \Cref{fig:pic-AL-3636};
\item separated pairs of triangles: $(3^2,4,3,4)$ lattice, \Cref{fig:AL-33434};
\item chains of triangles: $(3^3,4^2)$ lattice, \Cref{fig:AL-33344};
\item and segments of chains of triangles aligned on honeycomb lattice as a back-bone: maple leaf---$(3^4,6)$ lattice, \Cref{fig:AL-33336}.
\end{itemize}
Additionally, we inspect the system evolution for the triangular $(3^6)$ lattice (\Cref{fig:pic-AL-333333}), diluted triangulations, and classical random graphs with mean node degree equal to $k=4$ and $k=5$, which also characterize the above-mentioned lattices. 
The characteristics of the Archimedean lattices discussed in this work are collected in \Cref{tab:lattices}.
Social noise is modeled by the temperature parameter $T$ in the heat-bath scheme.

To compare results of planar with non-planar lattices, we repeat the calculations for small cliques, i.e. fully connected networks, of $N$ nodes. In computational sociology, the structure of cliques is often used to explore properties of small groups \cite{Galam_2002,Merlone_2024}. The case of $N=3$ (a triad) belongs to the set of Archimedean lattices. We will show that the obtained $U(T)$ dependence is the same for separate triads and for all Archimedean networks, where $U(T)$ is the work function defined in \Cref{sec:Calculations}. The cases of $N=4$ and $N=5$ are not planar. 
Yet, if a network is composed of cliques sharing only corners of the cliques, the obtained $U(T)$ is just a sum of work functions of individual cliques.

Our calculations for the cliques are analytical, and related to thermal equilibrium. The calculations for planar lattices are performed both with asynchronous and synchronous
scheme. The former is known to reproduce properly the thermodynamic
properties of systems in thermal equilibrium. The latter is sociologically justified if we intend to mimic the dynamics of opinions gathered periodically in polls.
In this case, a respondent is not aware of the answers of other respondents
during the same round. Such procedure of data collection has been applied in Reference~\onlinecite{usdata}, to give an example. Synchronous simulations allow us to identify metastable minima of the work function, which are visible as a departure from the asynchronous case.

The paper is organized as follows: in \Cref{sec:Methods} we describe the methodology of the studies (including the creation of adjacency matrices for the Archimedean lattices mentioned above, the heat-bath algorithm and calculation of the work function), \Cref{sec:Results} contains results of theoretical calculations, presentation of the mathematical proof that structurally ordered phase is limited to $T=0$ on any planar lattice and Monte Carlo simulations, and \Cref{sec:Discussion} is devoted to discussion of the obtained results. 

The paper is finalized with \Cref{app:Shapes} showing lattices and adjacency matrices shapes. Fortran95 procedures for creating adjacency matrices $\mathbf{A}$ for the lattices considered here are provided in the Supplemental Material \cite{SM}.

%% ###############################################
%% ###############################################
\section{\label{sec:Methods}Methods}
%% ###############################################
%% ###############################################

We keep the simulation methodology as presented for diluted/densified triangulations and classical random graphs \cite{2005.11402,2106.03054,2206.14226}.
In this approach,
the energy of the system is given by
%% -----------------------------------------------
\begin{equation}
E=-\sum_i\sum_{j>i}\sum_{k>j} a_{ij}x_{ij} \cdot a_{jk}x_{jk} \cdot a_{ki}x_{ki}, 
\label{eq:H}
\end{equation}
%% -----------------------------------------------
where every link $x_{ij}$ represents friendly ($x_{ij}=+1$) or hostile ($x_{ij}=-1$) attitudes between actors $i$ and $j$, while (binary and symmetric) adjacency matrix $\mathbf{A}=[a_{ij}]$ elements,
%% -----------------------------------------------
\begin{equation}
a_{ij}=\begin{cases}
1 & \iff i \text{ and } j \text{ are connected},\\
0 & \iff \text{otherwise},
\end{cases}
\end{equation}
%% -----------------------------------------------
define geometry of the lattice.

The equivalents of the lattice shapes shown in \Cref{fig:AL} mapped onto distorted triangular lattices embedded in square lattices and the corresponding adjacency matrices are presented in \Cref{app:Shapes}, \Cref{fig:kagome,fig:al33434,fig:al33344,fig:maple-leaf,fig:triangular}. 

The Heider balance can be easily identified by checking the system energy \cite{Antal_2005,Krawczyk_2017} per number of triangles $\Delta$ on the lattice,
%% -----------------------------------------------------
\begin{equation}
\label{eq:U}
E/\Delta\equiv-\frac{\Tr[(\mathbf{A}\circ\mathbf{X})^3]}{\Tr(\mathbf{A}^3)},
\end{equation}
%% -----------------------------------------------------
where $\circ$ stands for the Hadamard product of the matrices and the matrix $\mathbf{X}=[x_{ij}]$.
The system energy per triangle $E/\Delta$ is equal to $-1$ if and only if all triangles in the system are balanced.

%% -----------------------------------------------------
\begin{table*}
\caption{\label{tab:lattices}Characteristics of some Archimedean lattices. The second column indicates the lattice connectivity $k$. References to figures showing lattices' shapes (and their shapes embedded in a distorted triangular lattice) are given in the third and the fourth columns. Figures showing the shapes of the adjacency matrix $\mathbf{A}$ are indicated in the fifth column. In the sixth column the labels of listings supplying $\mathbf{A}$ are listed.
The latter are available in the Supplemental Material \cite{SM}}
\centering
\begin{ruledtabular}
\begin{tabular}{lrrrrr}
lattice & $k$ & shape & shape in \Cref{app:Shapes} & matrix $\mathbf{A}$ in \Cref{app:Shapes} & subroutine in \cite{SM}\\
\hline
$(3^6)$       & 6 & \Cref{fig:pic-AL-333333} & \Cref{fig:triangular-shape} & \Cref{fig:triangular-adjacency} & Listing~2\\
$(3^2,4,3,4)$ & 5 & \Cref{fig:AL-33434} & \Cref{fig:al33434-shape} & \Cref{fig:al33434-adjacency} & Listing~3\\
$(3^3,4^2)$   & 5 & \Cref{fig:AL-33344} & \Cref{fig:al33344-shape} & \Cref{fig:al33344-adjacency} & Listing~4\\
$(3^4,6)$     & 5 & \Cref{fig:AL-33336} & \Cref{fig:maple-leaf-shape} & \Cref{fig:maple-leaf-adjacency} & Listing~5\\
$(3,6,3,6)$   & 4 & \Cref{fig:pic-AL-3636} & \Cref{fig:kagome-shape}  & \Cref{fig:kagome-adjacency} & Listing~6\\
\end{tabular}
\end{ruledtabular}
\end{table*}
%% -----------------------------------------------------

The social noise level is mimicked by the temperature $T$ introduced in the probabilities of setting a hostile or friendly relation between actors.
If $a_{ij}\ne 0$ then evolution of the link between nodes $i$ and $j$ of value $x_{ij}$ is given by the heat-bath algorithm,
%% -----------------------------------------------
\begin{equation}
\label{eq:evol}
x_{ij}(t+1)=
    \begin{cases}
	+1 & \text{ with probability }p_{ij}(t),\\
	-1 & \text{ with probability }[1-p_{ij}(t)],
    \end{cases}
\end{equation}
%% -----------------------------------------------
where
%% -----------------------------------------------
\begin{equation}
\label{eq:evol_p}
    p_{ij}(t)=\frac{\exp[\beta\xi_{ij}(t)]}{\exp[\beta\xi_{ij}(t)]+\exp[-\beta\xi_{ij}(t)]}
\end{equation}
%% -----------------------------------------------
with $\beta=1/T$, 
and
%% -----------------------------------------------
\begin{equation}
\label{eq:evol_xi}
	\xi_{ij}(t)=\sum_{k} a_{ik} x_{ik}(t) \cdot a_{kj} x_{kj}(t).
\end{equation}
%% -----------------------------------------------

An ergodic Markov chain based on repeating local steps done with the heat-bath transition
probability
%% -----------------------------------------------
\begin{equation}
    p(A\rightarrow B) = \frac{e^{-\beta (E_B - E_A)}}{\sum_C e^{-\beta E_C}}
\end{equation}
%% -----------------------------------------------
leads to a stationary state with the partition function
%% -----------------------------------------------
\begin{equation}
    Z = \sum_{C} e^{-\beta E_C},
    \label{pf}
\end{equation}
%% -----------------------------------------------
where $C$ denotes all microscopic states.

Knowing the partition function $Z$ we can determine the mean value of energy of the system as
%% ---------------------------------------------------------------
\begin{equation}
\label{eq:UzZ}
U=-\frac{d}{d\beta} \ln Z.
\end{equation}
%% ---------------------------------------------------------------

%% ###############################################
%% ###############################################
\section{\label{sec:Results}Results}
%% ###############################################
%% ###############################################

%% ===============================================
\subsection{\label{sec:Calculations}Calculations}
%% ===============================================

For one triad, we have $2^3$ microscopic states $C$, four of them
with energy $E=-1 $ and four with energy $E=+1$; hence the partition function is 
%% ---------------------------------------------------------------
\begin{equation}
Z_1=\sum_{C}\exp(-\beta E_{C})=8\cosh(\beta),
\end{equation}
%% ---------------------------------------------------------------
and the mean value of energy is
%% ---------------------------------------------------------------
\begin{equation}
\label{eq:U1}
U_1=-\frac{d}{d\beta} \ln Z_1 =-\tanh(\beta).
\end{equation}
%% ---------------------------------------------------------------

For two triads with a common link, we have $2^5$ microscopic states. The energy spectrum contains eight states of energy equal $-2$, eight states of energy equal $+2$, and sixteen states of energy equal zero.
The direct summation over these states gives the partition function $Z_2 \propto Z_1^2$. Accordingly, 
%% ---------------------------------------------------------------
\begin{equation}
U_2=2U_1
\end{equation}
%% ---------------------------------------------------------------
and 
%% ---------------------------------------------------------------
\begin{equation}
\label{eq:Upair}
U_2/\triangle=-\tanh(\beta).
\end{equation}
%% ---------------------------------------------------------------

%% ---------------------------------------------------------------
\begin{figure}[bp]
\begin{center}
\includegraphics[width=0.99\columnwidth]{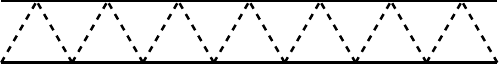}
\end{center}
\caption{\label{fig:chain}Chain of triangles appearing in $(3^3,4^2)$ lattice}
\end{figure}
%% ---------------------------------------------------------------

For a chain of triangles shown in \Cref{fig:chain}, we can distinguish two kinds of links: internal (dashed) and closing (solid) ones. For each configuration of internal links, two states of each triad are possible, with opposing energies of this triad: $-1$ for a balanced triad, and $+1$ for an imbalanced triad. The energy of the triad is determined by the state of the third, closing link in this triad. Summarizing, the 
partition function of the chain includes:
\begin{itemize}[noitemsep]
\item all configurations of internal links;
\item two states of each closing link, with opposite energies, for each triad.
\end{itemize}
In this sense, the energies of the configurations depend only on the states of closing links,
%% ---------------------------------------------------------------
\begin{equation}
  Z_{\Delta}=\sum_{i}\sum_{c}\exp(-\beta E(i,c)) =2^{\Delta+1}[2\cosh(\beta)]^\Delta,
\end{equation}
%% ---------------------------------------------------------------
where $\Delta$ is the number of triads, while $i$ and $c$ are related to the states of internal and closing links, respectively. As a consequence, the mean energy per one triangle is equal to
%% ---------------------------------------------------------------
\begin{equation}
\label{eq:Uchain}
U_{\text{chain}}/\triangle=-\tanh(\beta)
\end{equation}
%% ---------------------------------------------------------------
again.

Now let us consider a network of tetrahedrons, formed in a way that neighboring tetrahedrons have only common corners. An example of this structure in crystallography is YMn$_2$ \cite{Szytula_2004}. It is clear that states of links of different tetrahedra do not depend on each other. Therefore the system is equivalent to a set of separate tetrahedra. A direct enumeration of $2^6=64$ states of one tetrahedron gives the partition function
\begin{equation}
Z_{\text{tetrahedra}}=8e^{-4\beta}+48+8e^{4\beta},
\end{equation}
which gives the expression per one triad
\begin{equation}
\label{eq:U4}
U_{\text{tetrahedra}}(\beta)/\triangle=-\frac{e^{4\beta}-e^{-4\beta}}{e^{4\beta}+6+e^{-4\beta}}.
\end{equation}

\begin{figure}[bp]
\begin{center}
\includegraphics[width=0.99\columnwidth]{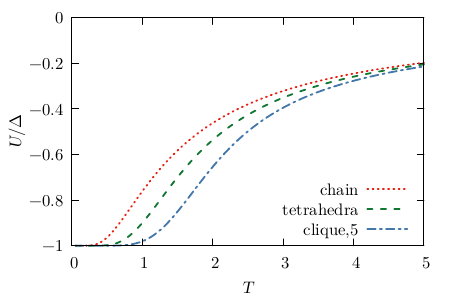}
\end{center}
\caption{\label{fig:UvsT-theory}The plots $U(T)$ for the chain of triangles, for cliques of four nodes (tetrahedra) and for cliques of five nodes, obtained from 
\Cref{eq:Uchain,eq:U4,eq:U5}}
\end{figure}

With some patience, the same procedure can be applied for networks of cliques of five or more nodes. For five nodes in a clique we get the work function per one triad
\begin{equation}
\label{eq:U5}
\begin{split}
U_{\text{clique,5}}(\beta)/\triangle=\\
-\frac{z^{10}+4z^4+3z^2-3z^{-2}-4z^{-4}-z^{-10}}{z^{10}+10z^4+15z^2+12+15z^{-2}+10z^{-4}+z^{-10}},
\end{split}
\end{equation}
where $z=e^{\beta}$.
A comparison of functions $U(T)$ calculated from \Cref{eq:Uchain,eq:U4,eq:U5} is shown in \Cref{fig:UvsT-theory}.

The calculated dependencies of $U(T)$ (per triangle) for a single triangle \eqref{eq:U1}, pairs of triangles \eqref{eq:Upair}, and the chain of triangles \eqref{eq:Uchain} are the same as for a one-dimensional Ising model with the nearest-neighbor interactions.
As we show below, this is not only a coincidence, but the Heider model (\ref{eq:H}) on a two-dimensional lattice with the asynchronous updates is equivalent to a one-dimensional Ising model with the nearest-neighbor interactions \cite{Lenz1920,Ising1925}.

%% ---------------------------------------------------------------
%% ---------------------------------------------------------------

\begin{figure}[htbp]
\includegraphics[width=0.99\columnwidth]{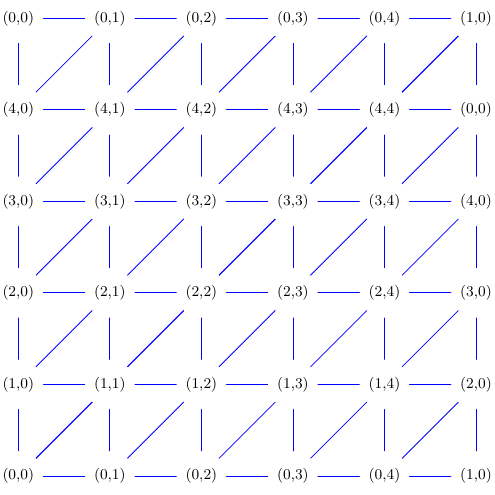}
\caption{\label{fig:triangultoroidal}An example of a triangular lattice with a toroidal topology with  twisted boundary conditions.
The lattice has $N=25$ nodes, indexed by $(i,j)$, with $i=0,\ldots,4$ and $j=0,\ldots,4$.
It has $\Delta=50$ triangles and $L=75$ links. The twisted boundary condition is encoded in a shift by one in the coordinate $j$ between the first and the last columns}
\end{figure}

Let us apply the following transformation, where $s_i=-1$, 
%% -----------------------------------------------
\begin{equation}
    x'_{ij}   = s_i x_{ij},     
    \label{gauge}
\end{equation}
%% -----------------------------------------------
to all links incident with the node $i$. The total energy of the system
\eqref{eq:H} remains the same $E_{A'}=E_A$. The gauge transformation \eqref{gauge} leaves the spin products $x_{ij}x_{jk}\ldots x_{si}$ unchanged in any closed loop.
The transformation \eqref{gauge} can be repeated for any node of the lattice. The configurations obtained in this way form an equivalence class with the same energy. Let $N$ be the number of nodes. It is easy to see that each equivalence class contains $2^{N-1}$  configurations since one can apply the transformation \eqref{gauge} to $(N-1)$ nodes. Applying it to  the $N$-th node would undo all changes as each spin $x_{ij}$ would be multiplied twice $s_i x_{ij} s_j$. Degeneracy can be removed by fixing values $x_{ij}$ on any spanning tree of the lattice, for example by assigning all edges of the tree the value one: $x_{ij}=1$. 
The spanning tree has $N-1$ edges, so fixing the gauge completely eliminates degeneracy.
Once the gauge is fixed, the partition function \eqref{pf} can be reduced to 
%% -----------------------------------------------
\begin{equation}
    Z = 2^{N-1} \sum_{a} e^{-\beta E_a},
\end{equation}
%% -----------------------------------------------
where $a$ are configurations of $x_{ij}$ for all edges except those which were fixed to one on the spanning tree.  We will illustrate this equivalence in detail on a triangular lattice with twisted boundary conditions,
as shown in \Cref{fig:triangultoroidal}, but it can be easily modified for any two-dimensional lattice.

%% ===============================================
%% ===============================================

\begin{figure}[htbp]
\includegraphics[width=0.99\columnwidth]{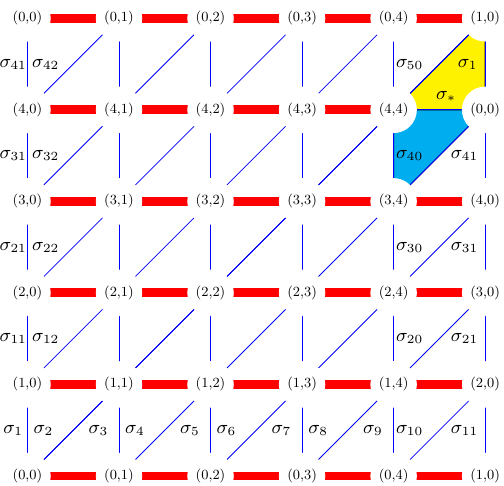}
\caption{\label{fig:gaugefixing}Gauge fixing: for all red edges $x_{ij}=1$. 
Any white triangle has two free edges, for which $x_{ij}=\pm 1$. 
There are only two triangles with three free edges.
They are drawn in yellow and turquoise}
\end{figure}

Let us introduce a gauge fixing as shown in \Cref{fig:gaugefixing}: $x_{ij}=1$ for all edges $ij$ on the line drawn in red.

%% ===============================================
%% ===============================================

Let us label the values of spins for the free links (which do not belong to the gauge tree) as $\sigma_1, \ldots, \sigma_{50}$ as shown in Fig.~\ref{fig:gaugefixing}, and denote the value of $\sigma$ on the link $(4,4)-(0,0)$ by $\sigma_*$.
We can write the partition function \eqref{pf} for this gauge fixing as
%% -----------------------------------------------
\begin{equation}
    Z = 2^{N-1}(Z_+ + Z_-),
\end{equation}
%% -----------------------------------------------
where $Z_+$ and $Z_-$ are sums over configurations with $\sigma_*=+1$ and $\sigma_*=-1$, respectively. 
It is easy to see that
%% -----------------------------------------------
\begin{equation}
    Z_+= \sum_{\sigma} e^{\beta \sum_{i=1}^{50} \sigma_i \sigma_{i+1}},
\end{equation}
%% -----------------------------------------------
where $\sigma_{51}=\sigma_1$. This is a partition function of the one-dimensional nearest-neighbor 
Ising model with periodic boundary conditions. For $\sigma_*=-1$, we have
%% -----------------------------------------------
\begin{equation}
    Z_-=\sum_{\sigma} e^{\beta (\sum_{i=1}^{50} \sigma_i \sigma_{i+1} - E)}, 
\end{equation}
%% -----------------------------------------------
where $E=2\sigma_{39}\sigma_{40} + 2\sigma_{50}\sigma_1$. Up to the last term $E$, $Z_-$ is
also equal to the partition function of the one-dimensional Ising model. This term introduces
some finite size corrections which can be neglected when the size $N = n\times n$ of the lattice increases with $n\to\infty$, so we see that for large lattices the original model is mapped into a one-dimensional Ising model on a chain of length $2n$. The mean energy per node for the Ising chain is known to be $-\tanh\beta$, so therefore in the Heider model with the asynchronous update scheme, the mean energy per triangle should
reproduce this result $U/\Delta\to -\tanh(\beta)$ in the limit $N\to\infty$.

%% ===============================================
\subsection{\label{sec:Computations}Computations}
%% ===============================================

For all the considered lattices, the initial state of each of the links $x_{ij}$ is chosen randomly to be $+1$ or $-1$ with equal probabilities, so that $U/\Delta \approx 0$ at time $t=0$.
Then, one of the two update procedures, synchronous or asynchronous, is applied for $t_{\text{max}}$ time steps, as given by~\Cref{eq:evol}.
In the case of the synchronous updating scheme, in one time step all links undergo this procedure simultaneously so that a new matrix $\mathbf{X}$ is created, which then replaces the old matrix before the next time step is performed. 
In the asynchronous approach, a single link is chosen randomly for an attempted update and its new value is determined instantly from~\Cref{eq:evol}; in one time step this procedure is repeated as many times as the total number of links, which means that statistically each link is taken into account once in a time step.
In both cases, the first 10\% of the performed time steps are not taken into account when calculating the results presented below, so that the system can relax from the initial state.
For most calculations $t_\text{max}=10^4$, but in some cases it is larger for $T<1$, as described below. 
In addition, all simulations are repeated 100 times and the mean value over those simulations is denoted by $\langle\cdots\rangle$.

%% ===============================================
\begin{figure*}[htbp]
%% -----------------------------------------------
\begin{subfigure}[b]{.85\columnwidth}
\caption{\label{fig:U_vs_T-s}}
\includegraphics[width=0.99\columnwidth]{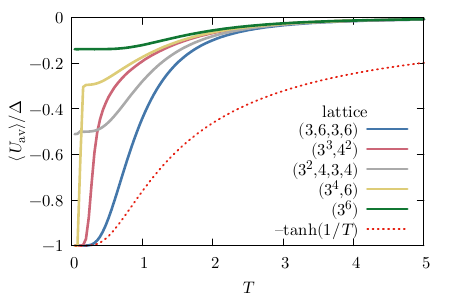}
\end{subfigure}
%% -----------------------------------------------
\begin{subfigure}[b]{.85\columnwidth}
\caption{\label{fig:U_vs_T-a}}
\includegraphics[width=0.99\columnwidth]{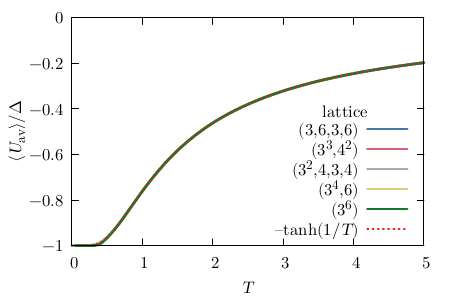}
\end{subfigure}
%% -----------------------------------------------
\begin{subfigure}[b]{.85\columnwidth}
\caption{\label{fig:U_vs_T-s-k4}}
\includegraphics[width=0.99\columnwidth]{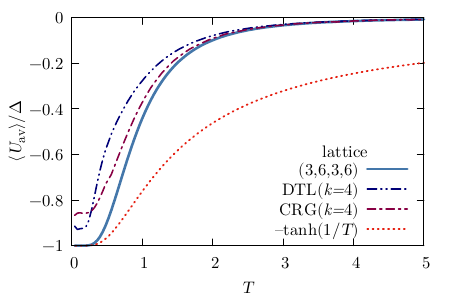}
\end{subfigure}
%% -----------------------------------------------
\begin{subfigure}[b]{.85\columnwidth}
\caption{\label{fig:U_vs_T-a-k4}}
\includegraphics[width=0.99\columnwidth]{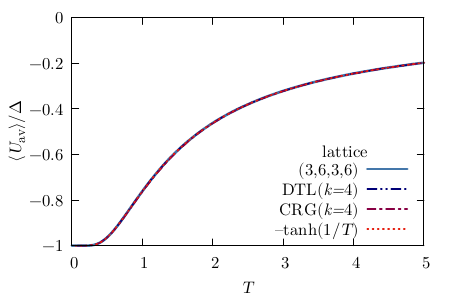}
\end{subfigure}
%% -----------------------------------------------
\begin{subfigure}[b]{.85\columnwidth}
\caption{\label{fig:U_vs_T-s-k5}}
\includegraphics[width=0.99\columnwidth]{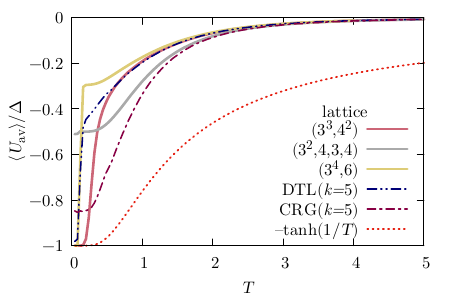}
\end{subfigure}
%% -----------------------------------------------
\begin{subfigure}[b]{.85\columnwidth}
\caption{\label{fig:U_vs_T-a-k5}}
\includegraphics[width=0.99\columnwidth]{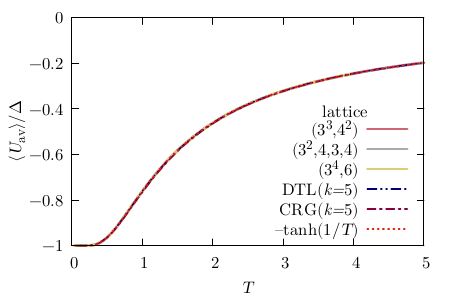}
\end{subfigure}
%% -----------------------------------------------
\begin{subfigure}[b]{.85\columnwidth}
\caption{\label{fig:U_vs_T-s-k6}}
\includegraphics[width=0.99\columnwidth]{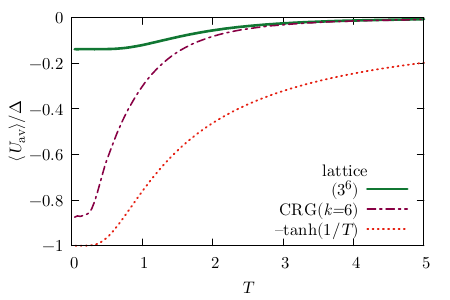}
\end{subfigure}
%% -----------------------------------------------
\begin{subfigure}[b]{.85\columnwidth}
\caption{\label{fig:U_vs_T-a-k6}}
\includegraphics[width=0.99\columnwidth]{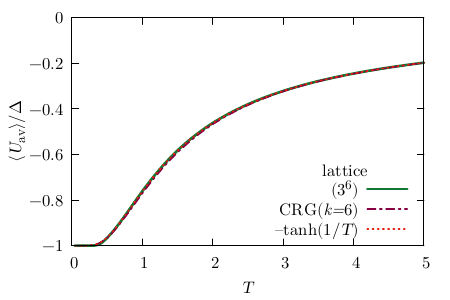}
\end{subfigure}
%% -----------------------------------------------
\caption{\label{fig:U_vs_T} $\langle U_\text{av}\rangle/\Delta$ for
synchronous 
(\subref{fig:U_vs_T-s}, \subref{fig:U_vs_T-s-k4}, \subref{fig:U_vs_T-s-k5}, \subref{fig:U_vs_T-s-k6}),
and asynchronous
(\subref{fig:U_vs_T-a}, \subref{fig:U_vs_T-a-k4}, \subref{fig:U_vs_T-a-k5}, \subref{fig:U_vs_T-a-k6}) update scheme. 
Nodes degree (on average, for the diluted triangular lattice and classical random graphs) $k=4$ (\subref{fig:U_vs_T-s-k4}, \subref{fig:U_vs_T-a-k4}),
$k=5$ (\subref{fig:U_vs_T-s-k5}, \subref{fig:U_vs_T-a-k5}), $k=6$ (\subref{fig:U_vs_T-s-k6}, \subref{fig:U_vs_T-a-k6})}
\end{figure*}
%% ===============================================

The mean energy per triangle defined in \Cref{eq:U} is therefore calculated in the final 90\% of the time steps and then averaged to find $U_\text{av}$, and then averaged again over all the simulations performed.
The results, denoted as $\langle U_\text{av}\rangle/\Delta$ are presented in~\Cref{{fig:U_vs_T}} as a function of temperature $T$, together with the curve given by
%% ===============================================
\begin{equation}
\label{eq:U_vs_T}
U/\Delta=-\tanh(1/T),
\end{equation}
%% ===============================================
corresponding to the theoretical results of \Cref{sec:Calculations}.
For all simulations on $(3^2,4,3,4)$ and $(3^3,4^2)$ lattices we assume $N=64$ which corresponds to $\Delta=64$; on $(3^4,5)$ $N=49$, $\Delta=56$; on the triangular lattice $(3^6)$ $N=49$, $\Delta=98$; and on $(3,6,3,6)$ $N=100$ with $\Delta=50$.

For comparison, we match the Archimedean lattices with the corresponding diluted triangular lattices (DTL)~\cite{2106.03054} and classical random graphs (CRG)~\cite{2206.14226} with $N=64$ and having the same average node degree $k = 4$, $5$, or $6$ as the connectivity of the considered Archimedean lattices.
The only Archimedean lattice discussed here with $k=4$ is $(3,6,3,6)$ and for synchronous updating DTL and CRG of the same $k$ reveal similar but not identical characteristics, with non-perfect balance $\langle U_\text{av}(T)\rangle/\Delta \approx -0.9$ at $T \to 0$ as visible in \Cref{fig:U_vs_T-s-k4}.
It should be noted, though, that the random nature of the DTL and CRG graphs means that the simulations are performed on lattices with the same average node degree and number of nodes, but with different structure of connections (graph egdes) which results, among other things, in the number of triads in those lattices that is not constant in all simulations but found to be $\Delta=37\pm 3$ for DTL or $\Delta=11\pm 3$ for CRG, with $N=64$ in both cases when $k=4$.
For $k=5$ (with $\Delta=75\pm 3$ for DTL or $\Delta=20\pm 4$ for CRG, \Cref{fig:U_vs_T-s-k5}) and $k=6$ (with $\Delta=36\pm 6$ for CRG, \Cref{fig:U_vs_T-s-k6}) with synchronous updating the results for DTL and CRG are similar to those for $k=4$, which also means that they are not exactly in line with those for Archimedean lattices, in particular in the case of $(3^2,4,3,4)$ lattice where the balanced state is not observed even in the low-temperature limit.
At higher $T$, the results for all Archimedean, DTL and CRG lattices agree.
For asynchronous updating, the results obtained for DTL and CRG at $k=4,5,6$ and presented in \Cref{fig:U_vs_T-a-k4,fig:U_vs_T-a-k5,fig:U_vs_T-a-k6} do not differ from those of Archimedean lattices.

%% ===============================================
\begin{figure}[htbp]
%% -----------------------------------------------
\begin{subfigure}[b]{.99\columnwidth}
\caption{\label{fig:U-T--MAPLELEAF-SYNC}}
\includegraphics[width=0.99\columnwidth]{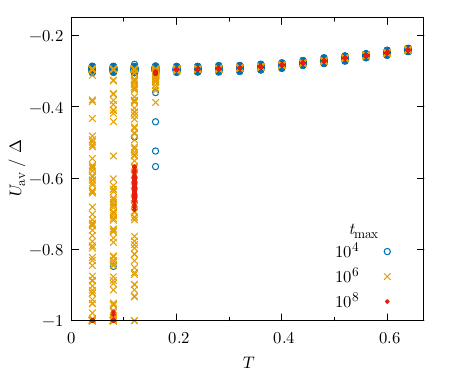}
\end{subfigure}
%% -----------------------------------------------
\begin{subfigure}[b]{.99\columnwidth}
\caption{\label{fig:U-T--SNUBSQUARE-SYNC}}
\includegraphics[width=0.99\columnwidth]{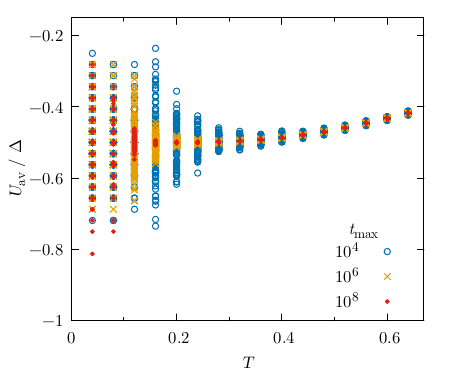}
\end{subfigure}
%% -----------------------------------------------
\caption{\label{fig:U-T--SYNC}
Values of $U_\text{av}/\Delta$ obtained in 100 simulations for each temperature $T$, 
for various simulation times $t_{\text{max}}=10^4$, $10^6$ and $10^8$, for (\subref{fig:U-T--MAPLELEAF-SYNC}) $(3^4,6)$ and (\subref{fig:U-T--SNUBSQUARE-SYNC}) $(3^2,4,3,4)$ lattice and synchronous scheme of link updates}
\end{figure}
%% ===============================================

The dispersion of the values obtained in 100 simulations is usually relatively small, especially when asynchronous updating is applied. 
In most cases, it is also true for synchronous updating, except for the $(3^2,4,3,4)$ and $(3^4,6)$ lattices at $T<1$ where it was required to perform $t_\text{max}=10^8$ time steps, and for DTL and CRG lattices where $t_\text{max}=10^6$ was used.
In particular in the case of $(3^4,6)$ lattice shorter simulations do not allow us to see that the balanced state is achieved at $T \to 0$, as visible in \Cref{fig:U-T--MAPLELEAF-SYNC} showing the values $U_\text{av}/\Delta$ obtained in all performed simulations.
\Cref{fig:U-T--SNUBSQUARE-SYNC} confirms that on the $(3^2,4,3,4)$ lattice the balanced state is not observed even in the low-temperature limit, and additionally the values of the work function obtained in the simulations performed are quantized and distributed over a relatively long interval.
The values of the work function observed at the lowest $T$ are spaced in the intervals of $2 / \Delta$, which corresponds to the difference related to the energetically nearest states.
The same quantization of the work function applies for any other planar lattice that contains triangles as it is associated with changing a single term ($x_{ij}x_{jk}x_{ik}$---where $i$, $j$, $k$ are triangle vertices) in sum in nominator of \Cref{eq:U} from $+1$ to $-1$ (or vice versa).

%% ===============================================
\begin{figure*}[htbp]
%% -----------------------------------------------
\begin{subfigure}[b]{.99\columnwidth}
\caption{\label{fig:TriadsPlus3}}
\includegraphics[width=0.99\columnwidth]{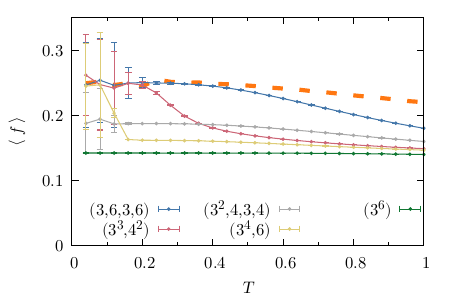}
\end{subfigure}
%% -----------------------------------------------
\begin{subfigure}[b]{.99\columnwidth}
\caption{\label{fig:TriadsPlus1}}
\includegraphics[width=0.99\columnwidth]{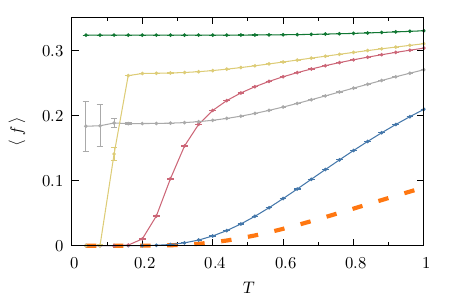}
\end{subfigure}
%% -----------------------------------------------
\begin{subfigure}[b]{.99\columnwidth}
\caption{\label{fig:TriadsMinus1}}
\includegraphics[width=0.99\columnwidth]{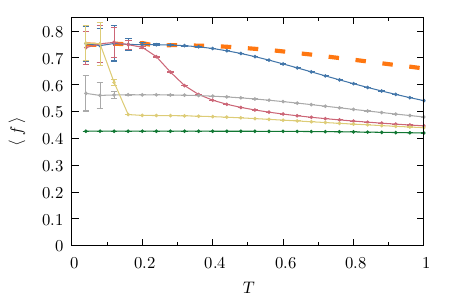}
\end{subfigure}
%% -----------------------------------------------
\begin{subfigure}[b]{.99\columnwidth}
\caption{\label{fig:TriadsMinus3}}
\includegraphics[width=0.99\columnwidth]{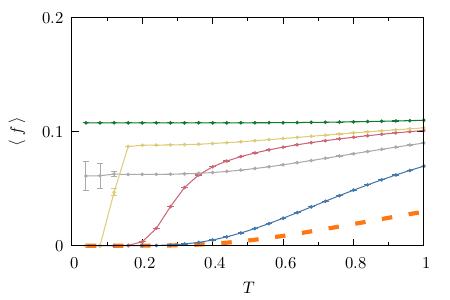}
\end{subfigure}
%% -----------------------------------------------
\caption{\label{fig:Triads_vs_T}Thermal evolution (in the low temperature limit $T<1$) of fractions of triads presented in \Cref{triangle} in the case of synchronous updating and different Archimedean lattices.
The panels (\subref{fig:TriadsPlus3}), (\subref{fig:TriadsPlus1}), (\subref{fig:TriadsMinus1}), and (\subref{fig:TriadsMinus3}) correspond to the triangles presented in Figures \ref{fig:tr-s+3}, \ref{fig:tr-s+1}, \ref{fig:tr-s-1}, and \ref{fig:tr-s-3}, respectively.
The error bars show the standard deviation of the results in 100 simulations.
Dashed line for asynchronous updating (common result for all lattices)
}
\end{figure*}
%% ===============================================

If any of the systems discussed is in the balanced state, it means that all triads are balanced, that is, they are in one of two possible configurations: with all positive links ($[+++]$, \Cref{fig:tr-s+3}) or with one positive and two negative links ($[+--]$, \Cref{fig:tr-s-1}).
The fraction of triads of a given kind, $\langle f \rangle$, as a function of $T$ is shown in \Cref{fig:Triads_vs_T}.
It reveals that in the cases where the balanced state is reached at $T \to 0$, about 25\% of the triads are in the $[+++]$ configuration, and the rest are in the $[+--]$ state. 
However, in that limit (only for $T<0.2$) those fractions vary substantially among the simulations, as shown by the error bars indicating the standard deviation, which is also the case for the $(3^2,4,3,4)$ lattice which does not reach the balanced state and for all triads in that case including $[++-]$ (\Cref{fig:tr-s+1}) and $[---]$ (\Cref{fig:tr-s-3}).
On the other hand, the $(3^6)$ triangular lattice produces almost identical results in all simulations.
In the balanced state observed in the low-temperature limit, the fractions of triads obtained using the asynchronous updating scheme (dashed lines in \Cref{fig:Triads_vs_T}) are the same as for the balanced state reached with synchronous updating.
Yet, in the case of asynchronous updating the increasing temperature causes much slower decrease of the number of balanced triads than for the synchronous approach, which is also visible in slower increase of the work function with increasing $T$.

%% ###############################################
%% ###############################################
\section{\label{sec:Discussion}Discussion}
%% ###############################################
%% ###############################################

We could expect that the results obtained with synchronous updating give systematically larger values of $U(T)$, than the thermodynamical counterparts. Further, and for the same reason, the frequencies of imbalanced triads obtained for the synchronous updating should exceed those obtained within the asynchronous scheme. The results from the synchronous updating are expected to depend on the length of simulation time because the system can escape from local minima of the work function after some characteristic time, which should depend on temperature. Our numerical results, shown in \Cref{fig:U-T--SYNC}, indicate that this effect appears indeed. Moreover, it is different in different Archimedean lattices. Apparently, the structure of local minima of the work function is sensitive to the lattice structure.

Accordingly, the synchronous updating for lower $T$ produces results that differ significantly from one Archimedean lattice to another.
\Cref{fig:U_vs_T-s} shows that the synchronous updating leads to $\langle U_\text{av}(T)\rangle/\Delta$ dependencies which are unique to each of the discussed Archimedean lattices. 
In the low-temperature limit, we observe perfectly balanced states, $\langle U_\text{av}\rangle/\Delta = -1$,  in the $(3,6,3,6)$, $(3^3,4^2)$ and $(3^4,6)$ lattices, but the decrease of energy with decreasing temperature occurs along a different curve for each of those lattices, and in all cases the values of the work function per triangle remain significantly larger than $-\tanh(1/T)$.
On the other hand, only a slight decrease in the work function is found for the triangular lattice $(3^6)$ and somewhat larger, to $\langle U_\text{av}(T\to 0)\rangle/\Delta \approx -0.5$, for the $(3^2,4,3,4)$ lattice.
For the asynchronous updating, the results in \Cref{fig:U_vs_T-a} prove that for all types of the considered Archimedean lattices the temperature dependence of the work function per triad is exactly the same and identical to the theoretically predicted $U/\Delta=-\tanh(1/T)$.

Besides local minima of $U$ and the related metastable (`jammed' \cite{Antal_2005}) states, the synchronous updating at $T=0$ admits limit cycles. In particular, limit cycles of length two, three and more have been identified in Reference~\onlinecite{PhysRevE.104.024307}. In the case of limit cycles of at least three steps, the detailed balance condition \cite[p. 406]{Reichl_1998} is violated. To see this, consider three states $A$, $B$ and $C$ in a limit cycle: $A$ is switched to $B$, $B$ to $C$ and $C$ to $A$. Their stationary probabilities are equal, but $B$ is never switched to $A$. For $T>0$, the transitions are not deterministic, yet the asymmetry of the rates of switching remains.
We admit, however, that for positive temperature,
the contribution of limit cycles to the system behavior is hard to be evaluated.

The equivalence of the planar Heider system and the one-dimensional Ising model, demonstrated at the end of \Cref{sec:Calculations}, allows us to deduce the lack of balance-imbalance phase transition at $T>0$. It is worthwhile to add that the consequence of this equivalence, i.e. the inequality $U(T)>-1$, seems to hold also for non-planar structures.  

As remarked in the Introduction, the synchronous updating scheme finds a counterpart in parallel decisions made by groups of voters. This scheme can also be used in settings of games played in parallel, as for example the Prisoner's Dilemma \cite{Hofbauer_Sigmund_1998}. There, the matrix of connections between players can be established by an appropriate structure of channels of exchange of information. Further, the incompleteness of information, characteristic for evolutionary games \cite{Hofbauer_Sigmund_1998}, can be modeled by the thermal noise. 

To summarize, the shape of the mean work function dependence on temperature for all planar networks investigated here, including the triangular lattice,  is the same as for a set of non-interacting triads. On the contrary, it appears different for a network of tetrahedrons, i.e. cliques of four nodes. The result for one tetrahedron is the same as for a network of tetrahedra, connected only by corners. Moreover, both above functions $U(T)$ are different from the one of a clique of five nodes. Finally, the latter formula for one clique is appropriate also for a network of such cliques, mutually connected by corners.

Concluding, the function $U(T)$ depends on the local topology of the network. All planar structures of triads, including the triangular lattice, display the same thermal dependence of $U$.  Further, an interaction of cliques through shared corners is not relevant for $U(T)$. On the contrary, the size of cliques matters. This raises an interesting question of whether and under which conditions the balance-imbalance transition is a collective effect.

\begin{acknowledgments}
The authors thank Zdzis{\l}aw Burda for a fruitful discussion of the gauge invariance theory applied to show that the perfectly balanced state is unreachable (\Cref{sec:Calculations}).
We gratefully acknowledge the Polish high-performance computing infrastructure PLGrid (HPC Center: ACK Cyfronet AGH) for providing computer facilities and support within computational grant no. PLG/2023/016259.
\end{acknowledgments}

%% #####################################################
\bibliography{heider,km,opiniondynamics,basics,ca,this}
%% #####################################################

\appendix

%% ###############################################
\section{\label{app:Shapes}Archimedean lattices mapped into square lattice and the shapes of their adjacency matrices}
%% ###############################################

\begin{figure*}[htbp]
\begin{subfigure}[b]{.65\columnwidth}
\caption{\label{fig:triangular-shape}}
\centering
\includegraphics[width=\textwidth]{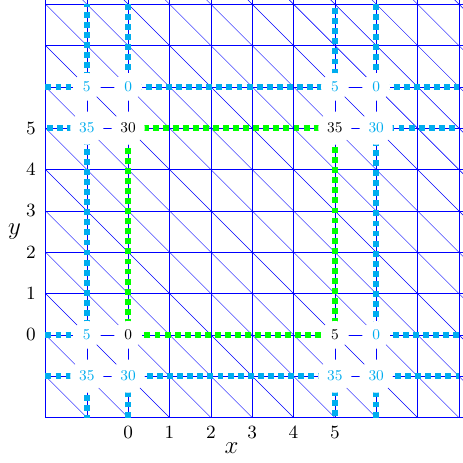}
\end{subfigure}
\hfill
\begin{subfigure}[b]{0.65\columnwidth}
\caption{\label{fig:triangular-adjacency}}
\centering
\includegraphics[width=\textwidth]{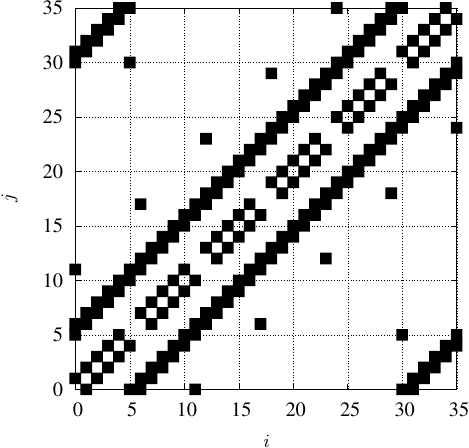}
\end{subfigure}
\caption{\label{fig:triangular}Triangular $(3^6)$ lattice, $k=6$, (\subref{fig:triangular-shape}) $N=W^2$ nodes at and inside green square are labeled as $i=0,\cdots,W^2-1$, $i=x+Wy$, $x,y\in\{0,1,2,\cdots,W-1\}$, $W=6$ (\subref{fig:triangular-adjacency}) and its adjacency matrix}
\end{figure*}

\begin{figure*}[htbp]
%% -----------------------------------------------
\begin{subfigure}[b]{.65\columnwidth}
\caption{\label{fig:al33434-shape}}
\includegraphics[width=\textwidth]{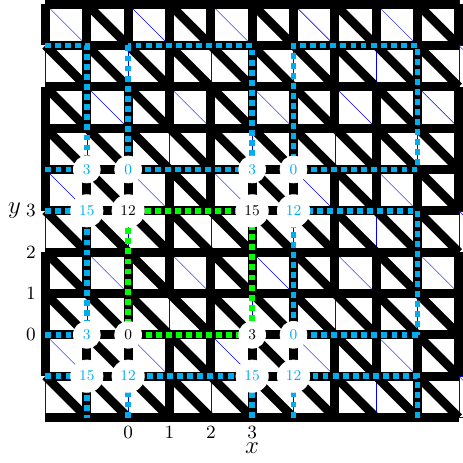}
\end{subfigure}
%% -----------------------------------------------
\hfill
%% -----------------------------------------------
\begin{subfigure}[b]{.65\columnwidth}
\caption{\label{fig:al33434-adjacency}}
\includegraphics[width=\textwidth]{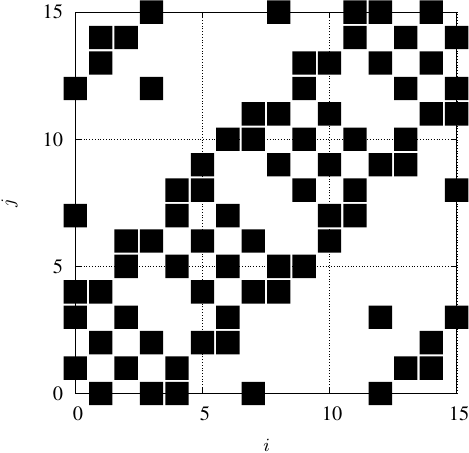}
\end{subfigure}
%% -----------------------------------------------
\caption{\label{fig:al33434}Archimedean $(3^2,4,3,4)$ lattice, $k=5$, (\subref{fig:al33434-shape}) $N=W^2$ nodes at and inside green square are labeled as $i=0,\cdots,W^2-1$, $i=x+Wy$, $x,y\in\{0,1,2,\cdots,W-1\}$, $W=4$ (\subref{fig:al33434-adjacency}) and its adjacency matrix}
\end{figure*}

\begin{figure*}[htbp]
\begin{subfigure}[b]{.65\columnwidth}
\caption{\label{fig:al33344-shape}}
\includegraphics[width=\textwidth]{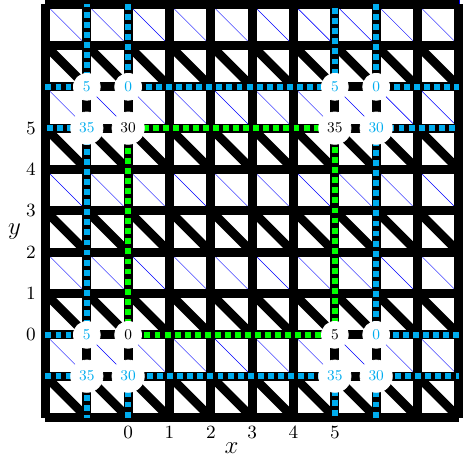}
\end{subfigure}
\hfill
\begin{subfigure}[b]{.65\columnwidth}
\caption{\label{fig:al33344-adjacency}}
\includegraphics[width=\textwidth]{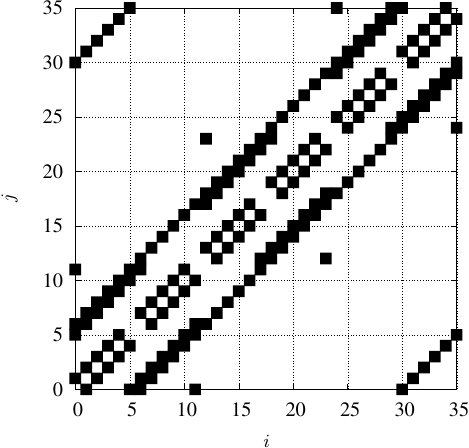}
\end{subfigure}
\caption{\label{fig:al33344}Archimedean $(3^3,4^2)$ lattice, $k=5$, (\subref{fig:al33344-shape}) $N=W^2$ nodes at and inside green square are labeled as $i=0,\cdots,W^2-1$, $i=x+Wy$, $x,y\in\{0,1,2,\cdots,W-1\}$, $W=6$ (\subref{fig:al33344-adjacency}) and its adjacency matrix}
\end{figure*}

\begin{figure*}[htbp]
%% -----------------------------------------------
\begin{subfigure}[b]{.69\columnwidth}
\caption{\label{fig:maple-leaf-shape}}
\includegraphics[width=\textwidth]{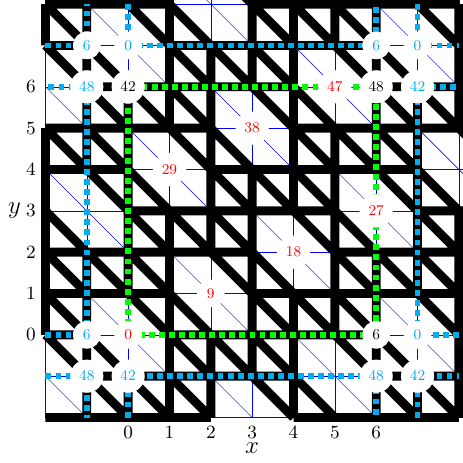}
\end{subfigure}
%% -----------------------------------------------
\hfill
%% -----------------------------------------------
\begin{subfigure}[b]{.69\columnwidth}
\caption{\label{fig:maple-leaf-adjacency}}
\includegraphics[width=\textwidth]{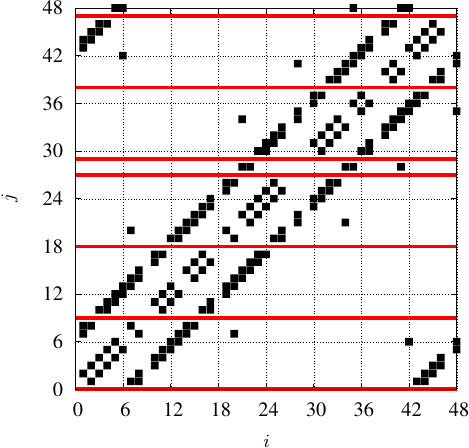}
\end{subfigure}
%% -----------------------------------------------
\caption{\label{fig:maple-leaf}Maple leaf $(3^4,6)$ lattice, $k=5$, (\subref{fig:maple-leaf-shape}) $N=W^2$ nodes at and inside green square are labeled as $i=0,\cdots,W^2-1$, $i=x+Wy$, $x,y\in\{0,1,2,\cdots,W-1\}$, $W=7$ (\subref{fig:maple-leaf-adjacency}) and its adjacency matrix}
\end{figure*}

\begin{figure*}[htbp]
\begin{subfigure}[b]{.69\columnwidth}
\caption{\label{fig:kagome-shape}}
\includegraphics[width=\textwidth]{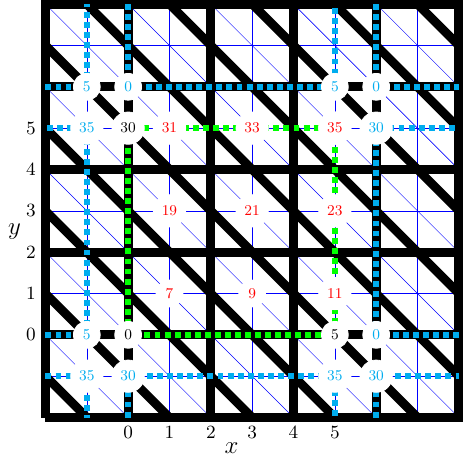}
\end{subfigure}
\hfill
\begin{subfigure}[b]{.69\columnwidth}
\caption{\label{fig:kagome-adjacency}}
\includegraphics[width=\textwidth]{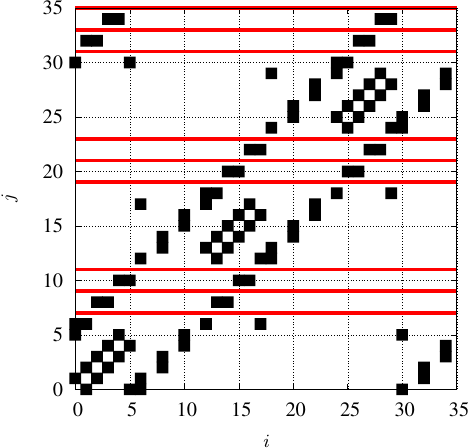}
\end{subfigure}
\caption{\label{fig:kagome}Archimedean $(3,6,3,6)$ lattice (kagom\'e lattice), $k=4$, (\subref{fig:kagome-shape}) $N=W^2$ nodes at and inside green square are labeled as $i=0,\cdots,W^2-1$, $i=x+Wy$, $x,y\in\{0,1,2,\cdots,W-1\}$, $W=6$ (\subref{fig:kagome-adjacency}) and its adjacency matrix}
\end{figure*}

In \Cref{fig:kagome-shape,fig:al33434-shape,fig:al33344-shape,fig:maple-leaf-shape,fig:triangular-shape} shapes of Archimedean lattices embedded in a distorted triangular lattice are presented for kagom\'e, $(3^2,4,3,4)$, $(3^3,4^2)$, maple-leaf, and triangular lattices, respectively.

In \Cref{fig:kagome-adjacency,fig:al33434-adjacency,fig:al33344-adjacency,fig:maple-leaf-adjacency,fig:triangular-adjacency} examples of adjacency matrices associated with lattices presented in \Cref{fig:kagome-shape,fig:al33434-shape,fig:al33344-shape,fig:maple-leaf-shape,fig:triangular-shape} are displayed.

%% ===============================================
%% ===============================================

%% ===============================================
%% ===============================================

%% ===============================================
%% ===============================================

%% ===============================================
%% ===============================================

%% ===============================================
%% ===============================================

\end{document}